\documentclass[twocolumn,prb,aps,showpacs]{revtex4}
\usepackage{textcomp,amssymb,graphicx}
\usepackage{hyperref}
\usepackage{amsbsy}
\usepackage{amsmath}
\usepackage{amssymb}
\usepackage[dvips]{color}
\usepackage{hyperref}
\usepackage{float}

\begin{document}
\title{Study of supersolidity in the two-dimensional Hubbard-Holstein model}
\author{A. Ghosh${^{1}}$}
\author{S. Kar${^{2}}$}
\author{S. Yarlagadda${^{1}}$}
\affiliation{${^1}$ CMP Division, Saha Institute of Nuclear Physics, HBNI, Kolkata, India}
\affiliation{${^2}$ Theoretical Physics Department, Indian Association for the Cultivation
of Science, Kolkata, India.}

\begin{abstract}
{
 We derive an effective Hamiltonian for the two-dimensional  Hubbard-Holstein model in the regimes of 
 strong electron-electron  
 and strong electron-phonon  interactions by using a nonperturbative approach. In the parameter region
 where the system manifests the existence of a correlated singlet phase, 
the effective Hamiltonian transforms to a $t_1-V_1-V_2-V_3$ Hamiltonian for  hard-core-bosons  on a checkerboard lattice. 
We employ quantum Monte Carlo simulations, involving  stochastic-series-expansion  technique, to obtain the ground
state phase diagram. At  filling $1/8$, as the strength of off-site repulsion increases,  the system undergoes a first-order
transition from a superfluid to a diagonal striped solid with ordering wavevector $\vec{Q}=(\pi/4,3\pi/4)$ or $(\pi/4,5\pi/4)$. 
Unlike  the one-dimensional situation, our results in the two-dimensional case reveal a supersolid phase
(corresponding to the diagonal striped solid)
around  filling $1/8$ and at large off-site repulsions.
Furthermore, for small off-site repulsions, we witness a  valence bond solid  
at one-fourth filling and 
tiny phase-separated regions at slightly higher fillings.
}
\end{abstract}
\maketitle
\section{Introduction}
Study of exotic quantum phases generated due to the coexistence or competition between diagonal and off-diagonal long range 
orders 
is a key area of continued interest in the
condensed matter community. In particular,  lattice supersolidity, which is the homogeneous coexistence of 
superfluidity/superconductivity and crystalline order in discrete lattices, has attracted considerable attention for  
 more than a few decades. In fact, lattice supersolidity has been observed in a number of systems such as the three-dimensional 
doped barium bismuthate\cite{bismuthate2,bismuthate3}; quasi-two-dimensional dichalcogenides\cite{quasi_2d_1} and layered molecular
crystals\cite{quasi_2d_2}; and quasi-one-dimensional doped trichalcogenide $\mathrm{NbSe}_3$\cite{quasi_1d_1} and doped spin 
ladder $\mathrm{Sr}_{14}\mathrm{Cu}_{24}\mathrm{O}_4$\cite{quasi_1d_2,quasi_1d_3}.

Furthermore, studies of cold atoms in optical lattices 
\cite{coldatom1,coldatom2,coldatom3,coldatom4,coldatom5}
have  paved the way to realize the coexistence of long range orders in a controlled way.
Though  numerous manifestations of supersolidity have been reported theoretically in bosonic systems, in different
 lattice geometries and with various kinds 
of interactions\cite{scalettar1,scalettar2,boninsegni1,zoller,wessel1,pinaki,troyer1,kar,Datta,lv2,wessel2,kedar1,kedar2,
melko1,boninsegni2,melko2,troyer2,wessel3,ye,mishra,ghosh,nigel}, it is only recently that such supersolid phases were realized
experimentally [by R. Landig {\em et al.} \cite{expt_supersolid}]. Since then, there is an upsurge in  the experimental effort to
  realize supersolid phases by utilizing both short-range and long-range  interactions.

Usually, diverse interactions can enrich the quantum phase diagram of the system 
by producing
various competing/cooperating orders.
Specifically,  strong electron-electron (e-e) interactions as well as  strong electron-phonon (e-ph) 
interactions  generate a rich phase diagram in systems such as 
the cuprates\cite{cuprates1,cuprates2}, 
the manganites\cite{manganites1,manganites2,manganites3}, 
and the fullerides\cite{fullerides}. 
In these correlated systems, a variety of exotic phases, such as superconductivity, charge-density-wave 
(CDW), spin-density-wave (SDW), etc. are manifested as an outcome of the interplay between  e-e and e-ph interactions.

A typical and simple model, to study the combined effect of  strong e-e and e-ph interactions, is the well-known Hubbard-Holstein model represented by the following Hamiltonian:
\begin{align}
 H_{hh}=&-t\sum_{j,\delta,\sigma}c^\dagger_{j+\delta\sigma}c_{j\sigma}+\omega_0\sum_j a_j^\dagger a_j\nonumber\\
 &+g\omega_0\sum_{j\sigma}n_{j\sigma}(a_j+a_j^\dagger)+U\sum_j n_{j\uparrow}n_{j\downarrow}
\end{align}
where $c^\dagger_{j\sigma}$ ($c_{j\sigma}$) denotes the creation (destruction) operator for spin-$\sigma$ electrons at site $j$, 
$t$ is the  hopping integral,  and the number operator $n_{j\sigma}=c^\dagger_{j\sigma}c_{j\sigma}$. Furthermore, $a_j^\dagger(a_j)$ corresponds to the 
creation (destruction) operator of phonons at site $j$ with dispersionless phonon frequency $\omega_0$, $g$ denotes the strength of the 
electron-phonon interaction, $U$ is the onsite Coulomb repulsion between electrons, and  $\delta$ represents 
the nearest-neighbors (NN). 

The Hubbard-Holstein model has been studied extensively in one, two and infinite dimensions at various fillings by employing diverse approaches such as quantum Monte Carlo (QMC)\cite{QMC1,QMC2,QMC3,QMC4,QMC5,QMC6}, exact diagonalization\cite{exact1,exact2,exact3}, density matrix renormalization group (DMRG)\cite{DMRG1,DMRG2}, dynamical mean field theory (DMFT)\cite{DMFT1,DMFT2,DMFT3,DMFT4,DMFT5,DMFT6,DMFT7,DMFT8,DMFT9}, semi-analytical slave boson approximations\cite{slaveboson1,slaveboson2,slaveboson3,slaveboson4,slaveboson5}, variational methods based on Lang-Firsov transformation\cite{variational1,variational2}, large-N expansion\cite{largeN}, Gutzwiller approximation\cite{Gutzwiller1,Gutzwiller2}, cluster approximation\cite{cluster}, and {static-auxiliary-field approximation\cite{Pai,Pinaki}}.


In this paper, we follow the approach  discussed in Refs. \onlinecite{sahinur1} and \onlinecite{sahinur2} and study the
two-dimensional Hubbard-Holstein model. {In contrast to Ref.\onlinecite{Pai},} our approach involves the non-adiabatic regime (i.e., $t/\omega_0 \le 1$). Furthermore, we employ a controlled analytic treatment of  the strong coupling regimes for both the e-ph ($g>1$) and e-e interactions ($U/t>1$) and take into account the dynamical quantum phonons.
The effective 
Hamiltonian consists of two major competing interactions---antiferromagnetic interaction between NN spins which favors the 
formation of singlets and NN repulsion between electrons which encourages CDW formation. Now, Ref. \onlinecite{sahinur1} showed 
that the quarter-filled one-dimensional Hubbard-Holstein model manifests a correlated singlet phase over a range of $U/t$ values,
whereas Ref. \onlinecite{sahinur2} demonstrated that this phase occurs at other fillings as well. In this work we concentrate only
on the correlated singlet phase in the two-dimensional version of the Hubbard-Holstein model. On representing a singlet by a 
hard-core-boson (HCB) at its center, the system of singlets on a periodic square lattice transforms into a system of HCBs on a 
checkerboard lattice. Using  quantum Monte Carlo (QMC) simulation involving stochastic-series-expansion (SSE) method,  we study the system 
at various filling fractions.
Our results for HCBs, at  filling $1/8$, indicates CDW order and
 unlike its one-dimensional 
analogue, exhibits supersolidity around  filling $1/8$. 
We explain the mechanism responsible for the formation of  the CDW as well as the supersolid phase (on the vacancy side and the interstitial side of the CDW).
Furthermore, our study at quarter-filling reveals mutually-exclusive existence
of valence bond solid (VBS) and superfluid (SF) phase.

The paper is organized as follows. In Sec. \ref{Eff_Hamiltonian}, we derive the effective Hamiltonian  and discuss the various terms. 
In Sec. \ref{HCB_model}, we show that the Hamiltonian of singlets on a square lattice transforms into a Hamiltonian of HCBs on a checkerboard lattice.
Sec. \ref{Numerical} deals with the numerical procedure as well as the order parameters used in our study. Next, the results are discussed in Sec. \ref{Results} and finally, conclusions are presented in Sec. \ref{Conclusion}.
\section{Effective Hamiltonian}\label{Eff_Hamiltonian}
The first step towards obtaining an effective Hamiltonian is to carry out the Lang-Firsov (LF) transformation, $H_{hh}^{LF}=e^S H_{hh} e^{-S}$ where $S=-g\sum\limits_{j\sigma}n_{j\sigma}(a_j-a_j^\dagger)$ and get the transformed Hamiltonian to be
\begin{align}
H_{hh}^{LF}=&-t\sum_{j,\delta,\sigma}X^\dagger_{j+\delta}c^\dagger_{j+\delta\sigma}c_{j\sigma}X_j+\omega_0\sum_ja_j^\dagger a_j\nonumber\\
&+(U-2g^2\omega_0)\sum_j n_{j\uparrow} n_{j\downarrow}-g^2\omega_0\sum_j(n_{j\uparrow}+n_{j\downarrow}),
\end{align}
with $X_j=e^{g(a_j-a_j^\dagger)}$. In terms of the composite fermionic operator, $d^\dagger_{j\sigma}\equiv c^\dagger_{j\sigma} X_j^\dagger$, the LF transformed Hamiltonian can be expressed as
\begin{align}
H_{hh}^{LF}=&-t\sum_{j,\delta,\sigma}d^\dagger_{j+\delta\sigma}d_{j\sigma}+\omega_0\sum_ja_j^\dagger a_j\nonumber\\
&+U_{eff}\sum_j n_{j\uparrow}^d n_{j\downarrow}^d-g^2\omega_0\sum_j(n_{j\uparrow}^d+n_{j\downarrow}^d),\label{LF_Hamiltonian}
\end{align}
where $n_{j\sigma}^d=d^\dagger_{j\sigma}d_{j\sigma}$ and $U_{eff}=U-2g^2\omega_0$. Since, the last term represents a constant polaronic energy, we can drop it 
without affecting the physics of the system. This leaves us with the realization that Eqn. (\ref{LF_Hamiltonian})  essentially represents the Hubbard model for composite fermions where the Hubbard interaction is given by $U_{eff}=U-2g^2\omega_0$. In the limit of large $U_{eff}/t$, with the help of a standard canonical transformation, the effective Hamiltonian, upto second order in the small parameter $t/U_{eff}$, can be expressed as
\begin{align}
H_{t-J-t_3}=&P_s\Big[-t\sum_{j,\delta,\sigma}d^\dagger_{j+\delta\sigma}d_{j\sigma}+\omega_0\sum_ja_j^\dagger a_j\nonumber\\
&+\frac{J}{2}\sum_{j,\delta}\left(\vec{S_j}\cdot \vec{S}_{j+\delta}-\frac{n_j^d n_{j+\delta}^d}{4}\right)\nonumber\\
&+t_3\sum_{j,\delta\neq\delta^{\prime},\sigma}d^\dagger_{j\bar{\sigma}}d_{j+\delta\sigma} d^\dagger_{j+\delta^{\prime}\sigma}d_{j\bar{\sigma}}\nonumber\\
&-t_3\sum_{j,\delta\neq\delta^{\prime},\sigma}d^\dagger_{j\sigma}d_{j+\delta\sigma}d^\dagger_{j+\delta^{\prime}\bar{\sigma}}d_{j\bar{\sigma}}\Big]P_s , \label{H_tjt3}
\end{align}
with $n_j^d=n_{j\uparrow}^d+n_{j\downarrow}^d$, $J=\frac{4t^2}{U_{eff}}$ and $t_3=J/4$. In the above expression $\vec{S}_j$ represents the spin operator for a fermion at site $j$ and the operator $P_s$ projects out 
double occupancy of any site.

In terms of the original fermionic operator, the effective Hamiltonian can be separated into two terms: (i) an unperturbed electronic Hamiltonian $H_0$ and (ii) a perturbative term $H_1$ in terms of the composite fermions. Thus,
\begin{align}
 H_{t-J-t_3}=H_0+H_1,
\end{align}
where
\begin{align}
H_0=&-te^{-g^2}\sum_{j,\delta,\sigma}P_s\left(c^\dagger_{j+\delta\sigma}c_{j\sigma}\right)P_s+\omega_0\sum_ja_j^\dagger a_j\nonumber\\
&+\frac{J}{2}\sum_{j,\delta}P_s\left(\vec{S_j}\cdot \vec{S}_{j+\delta}-\frac{n_j n_{j+\delta}}{4}\right)P_s\nonumber\\
&+\frac{Je^{-g^2}}{4}\sum_{j,\delta\neq\delta^{\prime},\sigma}P_s\left(c^\dagger_{j\bar{\sigma}}c_{j+\delta\sigma} c^\dagger_{j+\delta^{\prime}\sigma}c_{j\bar{\sigma}}\right)P_s\nonumber\\
&-\frac{Je^{-g^2}}{4}\sum_{j,\delta\neq\delta^{\prime},\sigma}P_s\left(c^\dagger_{j\sigma}c_{j+\delta\sigma}c^\dagger_{j+\delta^{\prime}\bar{\sigma}}c_{j\bar{\sigma}}\right)P_s\label{H_0} ,
\end{align}
and
\begin{align}
 H_1=-te^{-g^2}\sum_{j,\delta,\sigma}P_s\left[c^\dagger_{j+\delta\sigma}c_{j\sigma}\left({Y_+^{j \dagger}} Y_-^j-1\right)\right]P_s\label{H_1}.
\end{align}
In the above expression, the electron-phonon interaction is depicted by $H_1$ only through the term $Y_{\pm}^j\equiv e^{\pm g(a_{j+\delta}-a_j)}$. One should note that since $J/4\ll t$, we have ignored the following terms in $H_1$:
\begin{align}
&\frac{Je^{-g^2}}{4}\sum_{j,\delta\neq\delta^{\prime},\sigma}P_s\left[c^\dagger_{j\bar{\sigma}}c_{j+\delta\sigma} c^\dagger_{j+\delta^{\prime}\sigma}c_{j\bar{\sigma}}\left({Z_{+}^{j \dagger}} Z_{-}^j-1\right)\right]P_s\nonumber\\
&-\frac{Je^{-g^2}}{4}\sum_{j,\delta\neq\delta^{\prime},\sigma}P_s\left[c^\dagger_{j\sigma}c_{j+\delta\sigma}c^\dagger_{j+\delta^{\prime}\bar{\sigma}}c_{j\bar{\sigma}}\left({Z_{+}^{j \dagger}} Z_{-}^j-1\right)\right]P_s ,
\end{align}
where $Z_{\pm}^j\equiv e^{\pm g(a_{j+\delta^{\prime}}-a_{j+\delta})}$.

Performing a second order perturbation theory that is similar to the one  outlined in Ref. \onlinecite{sahinur1}, the effective Hamiltonian is obtained to be
\begin{align}
 H_{hh}^{\rm eff}\cong &-t_{\rm eff}h_{t_1}+\frac{J}{2}h_S-Vh_{nn}-t_2 h_{\sigma\sigma}\nonumber\\
 &-\left(t_2+J_3\right) h_{\sigma\bar{\sigma}}+J_3 h^{\prime}_{\sigma\bar{\sigma}} , \label{H_eff}
\end{align}

where
\begin{align}
 h_{t_1}=&\sum_{j,\delta,\sigma}P_s\left(c^\dagger_{j+\delta\sigma}c_{j\sigma}\right)P_s ,\\
 h_S=&\sum_{j,\delta}P_s\left(\vec{S_j}\cdot \vec{S}_{j+\delta}-\frac{n_j n_{j+\delta}}{4}\right)P_s ,\\
 h_{nn}=&\sum_{j,\delta,\sigma} \left(1-n_{j+\delta\bar{\sigma}}\right) \left(1-n_{j\bar{\sigma}}\right) n_{j\sigma}\left(1-n_{j+\delta\sigma}\right) ,\\
 h_{\sigma\sigma}=&\sum_{j,\delta\neq\delta^\prime,\sigma} \left(1-n_{j+\delta\bar{\sigma}}\right) \left(1-n_{j\bar{\sigma}}\right) \left(1-n_{j+\delta^\prime\bar{\sigma}}\right)\nonumber\\
 &\times\left[c^\dagger_{j+\delta\sigma}\left(1-2n_{j\sigma}\right)c_{j+\delta^\prime\sigma}\right] ,\\
 h_{\sigma\bar{\sigma}}=&\sum_{j,\delta\neq\delta^\prime,\sigma} \left(1-n_{j+\delta\bar{\sigma}}\right) \left(1-n_{j+\delta^\prime\sigma}\right)\nonumber\\
 &\times\left[c^\dagger_{j\sigma}c_{j+\delta\sigma}c^\dagger_{j+\delta^\prime\bar{\sigma}}c_{j\bar{\sigma}}\right] ,\\
 \end{align}
 and
 \begin{align}
 h^{\prime}_{\sigma\bar{\sigma}}=&\sum_{j,\delta\neq\delta^\prime,\sigma} \left(1-n_{j+\delta\bar{\sigma}}\right) \left(1-n_{j\sigma}\right) \left(1-n_{j+\delta^\prime\bar{\sigma}}\right)\nonumber\\
 &\times\left[c^\dagger_{j\bar{\sigma}}c_{j+\delta\sigma}c^\dagger_{j+\delta^\prime\sigma}c_{j\bar{\sigma}}\right] .
\end{align}
\begin{figure}[t]
\includegraphics[width=1\linewidth,angle=0]{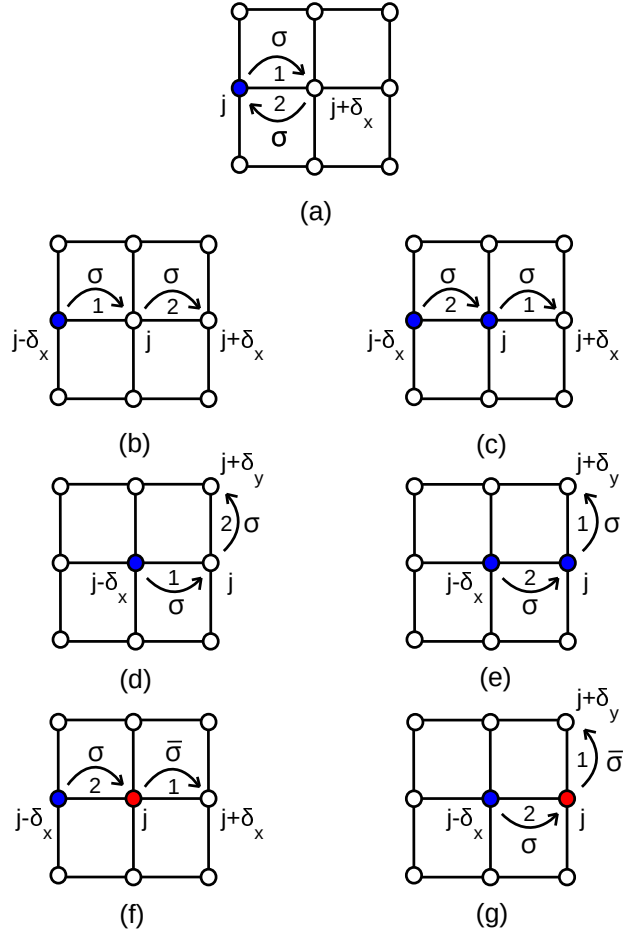}
\caption{(Color online) Different hopping processes which contribute to second-order perturbation theory: (a) $c^\dagger_{j\sigma}c_{j+\delta_x\sigma}c^\dagger_{j+\delta_x\sigma}c_{j\sigma}$, (b) $c^\dagger_{j+\delta_x\sigma} c_{j\sigma} c^\dagger_{j\sigma}c_{j-\delta_x\sigma}$, (c) $c^\dagger_{j\sigma} c_{j-\delta_x\sigma} c^\dagger_{j+\delta_x\sigma} c_{j\sigma}$, (d) $c^\dagger_{j+\delta_y\sigma} c_{j\sigma} c^\dagger_{j\sigma}c_{j-\delta_x\sigma}$, (e) $c^\dagger_{j\sigma} c_{j-\delta_x\sigma} c^\dagger_{j+\delta_y\sigma} c_{j\sigma}$, (f) $c^\dagger_{j\sigma} c_{j-\delta_x\sigma}c^\dagger_{j+\delta_x\bar{\sigma}}c_{j\bar{\sigma}}$, and (g) $c^\dagger_{j\sigma} c_{j-\delta_x\sigma}c^\dagger_{j+\delta_y\bar{\sigma}}c_{j\bar{\sigma}}$. Empty circles denote sites without electrons;  filled blue and red circles represent sites occupied by electrons with spin $\sigma$ and spin $\bar{\sigma}$ respectively.}
\label{hopping_process}
\end{figure}
The different coefficients for the various terms present in Eqn. (\ref{H_eff}) are defined as follows: $t_{\rm eff}=te^{-g^2}$, $J\equiv\frac{4t^2}{U-2g^2\omega_0}$, $V\simeq t^2/2g^2\omega_0$, $t_2\simeq t^2e^{-g^2}/g^2\omega_0$ and $J_3=Je^{-g^2}/4$. Out of the six terms of the effective Hamiltonian $H_{hh}^{\rm eff}$, four terms contribute to the kinetic energy of the system. However, due to the presence of $e^{-g^2}$ in the coefficients,
the contribution of the kinetic terms is small compared to that from the remaining two interaction terms.
The first contribution in the kinetic energy is from the NN hopping term $-t_{\rm eff}h_{t_1}$ in which the hopping coefficient is given by a reduced hopping integral $t_{\rm eff}=te^{-g^2}$. 
Next, the term $-t_2 h_{\sigma\sigma}$ is represented by the typical processes shown in Figs.  \ref{hopping_process}(b), \ref{hopping_process}(c)
\ref{hopping_process}(d) and \ref{hopping_process}(e). Figs.  \ref{hopping_process}(b) and \ref{hopping_process}(d) depict double hopping
of a single particle to next-to-next-nearest-neighbor (NNNN) site and 
next-nearest-neighbor (NNN) site, respectively; contrastingly, Figs.  \ref{hopping_process}(c) and \ref{hopping_process}(e) describe
processes where a pair of electrons of spin $\sigma$ hop sequentially along a straight path and a right-angled path, respectively.
\begin{figure}[t]
\includegraphics[width=0.8\linewidth,angle=0]{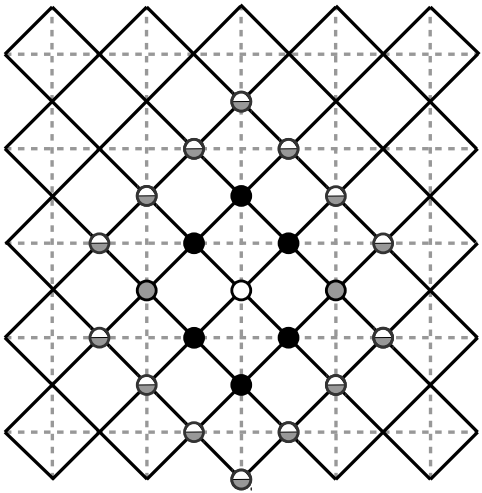}
\caption{(Color online) Checkerboard lattice constructed by joining the midpoints of the edges of a square lattice (indicated by the dashed lines). The filled black circles denote the six NN of the HCB depicted by the white circle, whereas the filled gray circles stand for 
next-nearest-neighbor (NNN) sites. The half-filled gray circles are next-to-next-nearest-neighbor (NNNN) sites for which the repulsion is half of the one felt for the filled gray sites.}
\label{checkerboard1}
\end{figure}
The next  term 
$-\left(t_2+J_3\right) h_{\sigma\bar{\sigma}}$
is represented by the typical hopping processes in  Figs. \ref{hopping_process}(f) and \ref{hopping_process}(g) which are similar to the hopping
processes shown in Figs. \ref{hopping_process}(c) and \ref{hopping_process}(e), respectively, but with the involved pair of electrons now having opposite spins $\sigma\bar{\sigma}$.
Lastly, the terms $ J_3 h^{\prime}_{\sigma\bar{\sigma}}$ 
implies NN spin-pair $\sigma\bar{\sigma}$ hopping similar to that depicted in Figs. \ref{hopping_process}(f) and \ref{hopping_process}(g), respectively, 
but with the spin-pair $\sigma\bar{\sigma}$ flipping to $\bar{\sigma}\sigma$. Thus, $h^{\prime}_{\sigma\bar{\sigma}}$ acting on a singlet state results in another singlet state displaced by one NN distance and with a negative sign. 

Now, the NN spin-spin interaction term $Jh_S$ and NN repulsion term $-Vh_{nn}$ dominate over the remaining hopping terms in the effective Hamiltonian. As discussed in the Refs. \onlinecite{sahinur1} and \onlinecite{sahinur2}, at larger $J$ values, a phase separated single cluster is formed because the spin-spin interaction dominates over the NN repulsion. 
As the $J/V$ value is decreased, the system undergoes a 
quantum phase transition to a correlated NN singlet phase where two NN particles pair to form a singlet\cite{hohenadler}. This correlated singlet phase persists over a range of $J/V$ values; at even smaller values,
a phase with separated spins is realized. It was also shown that the window of $J/V$, for which the correlated singlet phase exists, is broader for larger $g$ values. { Even for the case of the two-dimensional Hubbard-Holstein model, we expect similar results to hold and we present supporting arguments as follows.  
In the cluster regime, based on Monte Carlo simulation of a two-dimensional Heisenberg antiferromagnet\cite{swanson}, the energy/site 
 $= -0.672 J + 2(2V-\frac{J}{4})$. On the other hand, for  separated singlets in the correlated singlet
   phase the energy/site $= - 0.375 J + \frac{1}{2}(2V-\frac{J}{4})$. Thus the cluster phase prevails when
 $= -0.672 J + 2(2V-\frac{J}{4}) <   - 0.375 J + \frac{1}{2}(2V-\frac{J}{4})$ or equivalently, when $U < 3.792 g^2\omega_0$.
 Next, the transition from the correlated-singlet phase to the separated-spin phase occurs when singlets dissociate
 and is independent of the dimension of the system;
 this transition occurs when,  for the correlated singlet phase, the energy/site $- 0.375 J + \frac{1}{2}(2V-\frac{J}{4}) \approx 0$,
 i.e., $U \approx 6 g^2\omega_0$.}
 
In this work, we concentrate on the region of the parameter space where the correlated singlet phase is manifested.
\section{$t_1-V_1-V_2-V_3$ hard-core-boson model on a checkerboard lattice}\label{HCB_model}
In the correlated singlet phase, each NN singlet can be represented as a HCB located at the center of the singlet. 
Thus, the system of NN singlets on a periodic square lattice transforms into a system of HCBs on a checkerboard lattice;
the resulting checkerboard lattice
 is constructed by joining the midpoints of the edges of the underlying square lattice (see Fig. \ref{checkerboard1}). 
 Now, there are two processes by which the singlets can transport in the system. The first process corresponds to
 NN hopping of spin-pair $\sigma\bar{\sigma}$ and is  represented by $h_{\sigma\bar{\sigma}}$ (without spins flipping) and $h^{\prime}_{\sigma\bar{\sigma}}$ 
 (involving flipping the spins).
 The second process is a consequence of the presence of the NN hopping $h_{t_1}$ in $H_{hh}^{\rm eff}$; this is a second order 
 process which involves breaking of a bound singlet state (with binding energy $E_B=-J+2V=-J+t^2/g^2\omega_0$) and hopping of the constituent spins.
 Now, the spins can hop in two different ways: (a) each spin hops to its NN site sequentially [in a manner given by Figs. \ref{hopping_process}(f)
 and \ref{hopping_process}(g)]  
  and generating the corresponding term  $-t_b h_{\sigma\bar{\sigma}}$ with $t_b\equiv t^2e^{-2g^2}/|E_B|$; 
  and (b) any one of the two constituent spins hops to its NN site (along $x$ or $y$ directions) and comes back [yielding the corresponding term $-t_b h_{nn}$]. All these processes effectively describe the NN hopping ($t_1$) of the HCBs in the checkerboard lattice. For example,
  in Fig. \ref{checkerboard1}, a HCB residing at the site denoted by a white circle can hop to its six NN sites represented by the filled black circles.
  Now, no pair of singlets can share a common site. Therefore, the NN repulsion ($V_1$) between two HCBs in the checkerboard lattice is 
  essentially infinity. 
Next, the NN repulsion between two electrons in the square lattice (coming from the terms $Jh_S$ and $-Vh_{nn}$ in the expression of $H_{hh}^{\rm eff}$) 
gives rise to the NNN repulsion and the NNNN repulsion between two HCBs in the checkerboard lattice. To understand this, in Fig. \ref{checkerboard1}, 
consider two HCBs residing at the white and any one of the two filled gray sites. Corresponding to this situation, in the original square lattice there will be two pairs of electrons which are NN, thus increasing the energy of the system by an amount $2(2V-J/4)$. In other words, repulsion $V_2=2(2V-J/4)$ 
is felt between the HCBs residing at the white circle and its NNN sites denoted by filled gray circles. 
On the other hand, the repulsion felt between the white circle and its fourteen NNNN sites, depicted by the half-gray circles, is $V_3=V_2/2$.

Finally the effective Hamiltonian governing the HCBs in the checkerboard lattice is given by
\begin{align}
 H_b=&-t_1\sum_{\langle i,j\rangle}\left(b^\dagger_i b_j+{\rm H.c.}\right)+V_1\sum_{\langle i,j\rangle} n_i n_j\nonumber\\ &+V_2\sum_{\langle\langle i,j\rangle\rangle} n_i n_j +V_3\sum_{\langle\langle\langle i,j\rangle\rangle\rangle} n_i n_j ,
 \label{Hamiltonian_hcb}
\end{align}
where $b_j$ $(b_j^\dagger)$ denotes the destruction (creation) operator for a HCB at site $j$ with $n_j=b_j^\dagger b_j$ being the number operator. 
Here, the symbol $\langle i,j\rangle$ stands for a NN pair of sites, whereas $\langle\langle i,j\rangle\rangle$ and $\langle\langle\langle i,j\rangle\rangle\rangle$ represent NNN pair and NNNN pair, respectively. The coefficients of the different terms of $H_b$ are given as follows: $t_1=(t_2+2J_3+t_b)$, $V_1=\infty$, $V_2=2(2V-J/4)$ and $V_3=V_2/2$. 
\section{Numerical Calculations}\label{Numerical}
To study the system of HCBs in the checkerboard lattice depicted by Fig. \ref{checkerboard1}, we employ quantum Monte Carlo (QMC) simulation
involving  stochastic-series-expansion (SSE) technique \cite{sandvik_paper,sandvik_review} 
with directed loop updates\cite{dir_loop1,dir_loop2}. To achieve the above end, first we rewrite the Hamiltonian $H_b$ in terms of spin-$1/2$ operators by identifying $b_j^\dagger=S_j^+$, $b_j=S_j^-$ and $n_j=S_j^z+\frac{1}{2}$. We recast the effective Hamiltonian $H_b$ for HCBs as an extended ${\rm XXZ}$ spin-$1/2$ Hamiltonian, which, in units of $2t_1$, is given by
\begin{align}
 H=&\sum_{\langle i,j\rangle}\left[-\frac{1}{2}\left(S_i^+ S_j^- + {\rm H.c.}\right)+\Delta_1 S_i^z S_j^z\right]\nonumber\\
 &+\sum_{\langle\langle i,j\rangle\rangle}\Delta_2 S_i^z S_j^z +\sum_{\langle\langle\langle i,j\rangle\rangle\rangle}\Delta_3 S_i^z S_j^z-h\sum_i S_i^z ,
\end{align}
where $\Delta_1=V_1/2t_1$, $\Delta_2=V_2/2t_1$, and $\Delta_3=V_3/2t_1$.
Furthermore, we have introduced the variable $h$ (a dimensionless external magnetic field); upon tuning $h$,
 we can access different magnetizations (or filling-fractions) of the system.
%

Due to the presence of a hopping term in the Hamiltonian,  superfluidity is expected; on the other hand, large repulsions indicate 
the possibility of a CDW. Hence, to study the competition or coexistence of these two long-range orders, we choose two order parameters: structure factor $S(\vec{Q})$ (for diagonal long-range order) and superfluid density $\rho_s$ (for off-diagonal long-range order). The expression for the structure factor per site is given as
\begin{align}
 S(\vec{Q})=\frac{4}{N^2}\sum_{i,j} e^{i\vec{Q}\cdot(\vec{R_i}-\vec{R_j})}\langle S_i^z S_j^z\rangle
\end{align}
where $\langle\cdots\rangle$ represents ensemble average. We study $S(\vec{Q})$ for all possible values of $\vec{Q}$ and identify the ones that
produce peaks in the structure factor.

The superfluid density, in terms of the fluctuation of winding numbers, is expressed as
\begin{align}
\rho_s=\frac{1}{2\beta}\langle W_x^2+W_y^2\rangle
\end{align}
where $W_x$ and $W_y$ denote the winding numbers along $x$ and $y$ directions, respectively; $\beta$ is the inverse temperature. Furthermore, $W_x$ can be calculated from the total number of operators transporting spin in the positive and negative $x$ directions (i.e., $N_x^+$ and $N_x^-$)
using the expression $W_x=\frac{1}{L_x}(N_x^+-N_x^-)$, where $L_x$ is the linear dimension of the lattice along the $x$ direction.

Now,  due to particle-hole symmetry, Eqn. (\ref{Hamiltonian_hcb}) corresponds to HCB particles (holes) for particle density between $0$ and $1/2$ 
($1/2$ and $1$).
The NN repulsion between two HCB particles or holes [i.e., $V_1$ in Eqn. (\ref{Hamiltonian_hcb})] is infinity.
In the filling-fraction range $1/4$ and $1/2$ ($1/2$ and $3/4$), the HCB particles (holes)
cannot be arranged so that no two HCB particles (holes) are on neighboring sites. In other words, for fillings of HCB particles (holes)
between $1/4$ and $1/2$ ($1/2$ and $3/4$)
on a checkerboard lattice, which corresponds to fillings of electrons between $1/2$ and $1$ ($1$ and $3/2$) on a square lattice,
our theory of correlated singlet phase of electrons does not hold. Therefore, on a checkerboard lattice of HCBs, we restrict our interest to particle
fillings in the range $0$ and $1/4$ (or $3/4$ and $1$).
In this paper,
we vary the magnetization from $1/4$ to $1/2$ which means decreasing  the hole density from $1/4$ to $0$.
\begin{table}[t]
\begin{tabular}{|c|c|c|c|c|}
\hline
 $h$ & $70.0$ & $75.0$ & $85.25$ & $92.0$\\ \hline
 $\tau_{int}$ & $~~1154~~$ & $~~56182~~$ & $~~420361~~$ & $~~3747~~$\\ \hline
\end{tabular}
\caption{{Autocorrelation times calculated for $\Delta_1=16$, $\Delta_2=10$ and $\Delta_3=5$ with $\epsilon_1=8$, $\epsilon_2=10/4$ and $\epsilon_3=\epsilon_2/2$; the magnetic fields are chosen close to the transitions as well as far from the transitions (see Fig. \ref{field_vs} for details).}}\label{table1}
\end{table}

Next, in the presence of large anisotropy (i.e., large values of $\Delta_1$, $\Delta_2$ and $\Delta_3$) SSE suffers from significant slowing down. 
Therefore, due to numerical restrictions we cannot use the actual values of the longitudinal couplings; instead, we use large enough cutoff values 
so that the physics  remains unaltered. These cutoff values must be chosen keeping $\Delta_1$ sufficiently larger than the other $\Delta_i$'s 
so that the HCBs always avoid NN occupation. On the other hand, $\Delta_2$ and $\Delta_3$ must be large compared to $t_1$, 
but certainly smaller than $\Delta_1$. We will discuss the cutoff values for the longitudinal couplings in the next section.


As discussed in Ref. \onlinecite{Batrouni}, simulating at low enough temperatures such that $\beta\sim L$ with $L$ being the linear 
dimension of the $L\times L$ square lattice, we can capture the ground state properties of a system using SSE. Since the values of the measured observables were the same
(within the error bars of our calculations) for both $\beta=3L/2$ and $\beta=2L$, we report the results for $\beta=3L/2$ in our simulations.
It is worth mentioning here that in SSE a parameter $\epsilon_i$ is introduced to make the matrix elements positive\cite{kar}. This parameter is usually very small. However, in cases with large anisotropy,  value of $\epsilon_i$ can affect the autocorrelation times. In such cases, we need to use larger values of $\epsilon_i$ to take care of the autocorrelation times. To make sure that the bin size is always much larger than the autocorrelation times,
we calculate the autocorrelation time ($\tau_{\rm int}$) given by the following formula
\begin{align}
 \tau_{\rm int}[m]=\frac{1}{2}+\sum_{t=1}^\infty A_m(t)
\end{align}
where
\begin{align}
 A_m(t)=\frac{\langle m(i+t)m(i)\rangle-\langle m(i)\rangle^2}{\langle m(i)^2\rangle-\langle m(i)\rangle^2}
\end{align}
with $i$ and $t$ representing the Monte Carlo steps and $\langle\cdots\rangle$ the average over the time $i$.
Based on the autocorrelation times obtained, for $\Delta_1$ we use $\epsilon_1=\Delta_1/2$; whereas $\epsilon_2=\Delta_2/4$ ($\epsilon_3=\Delta_3/4$) is good enough to restrict the autocorrelation time within affordable limits. An estimate of the autocorrelation time for $\Delta_1=16$, $\Delta_2=10$ and $\Delta_3=5$ is given in Table \ref{table1}. The magnetic fields are chosen close to the transitions, where the autocorrelation time is expected to be larger, as well as away from them.
The bin size used for all numerical calculations is $16,00,000$ to make sure that the autocorrelation time is well within the bin size for all magnetic fields.
\begin{figure}[t]
\includegraphics[width=0.8\linewidth,angle=0]{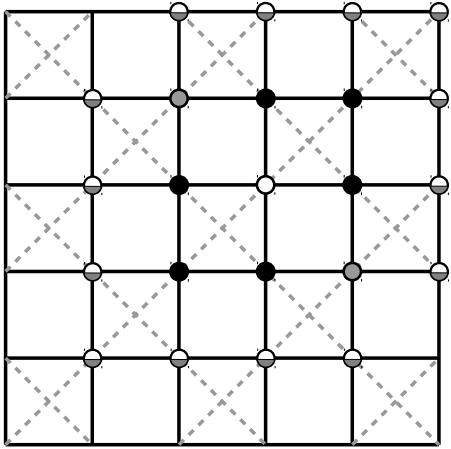}
\caption{(Color online) Checkerboard lattice of second type which is a part of the original checkerboard lattice (in Fig. \ref{checkerboard1}) and rotated by $45$\textdegree ~angle. The filled black circles denote the six NN of the HCB depicted by the white circle, whereas the filled gray circles stand for NNN sites. The half-filled gray circles are NNNN sites for which the repulsion is half of the one felt for the filled gray sites.}
\label{checkerboard2}
\end{figure}
\section{Results and discussions}\label{Results}
For numerical simulations, we can consider two types of lattices. 
A checkerboard lattice, constructed from an underlying $L\times L$ square lattice (see Fig. \ref{checkerboard1}), contains $2\times L\times L$ number of sites.
Alternately,  an $L\times L$  checkerboard lattice, as shown in
Fig. \ref{checkerboard2},  can be obtained via a  $45$\textdegree ~rotation
of the lattice of Fig. \ref{checkerboard1}.
In the thermodynamic limit,
either of the choices is supposed to yield the correct results; we have checked that even for a small system size with $L=8$, both the lattices produce the same results. 
Thus, at large anisotropies, simulation time can be lowered by considering a $L\times L$  checkerboard lattice.
In this paper, we present the results for HCBs on a $16\times 16$ checkerboard lattice of the second type, as depicted in Fig. \ref{checkerboard2}.

\begin{figure}[t]
\includegraphics[width=1.0\linewidth,angle=0]{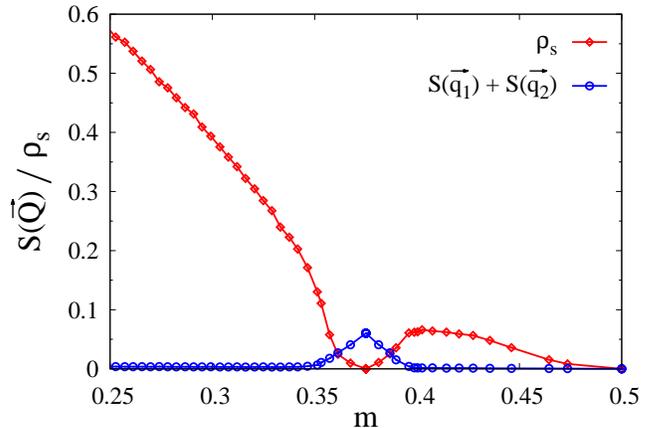}
\caption{(Color online) Plots of structure factor $S(\vec{Q})$ and superfluid density $\rho_s$ vs magnetization $m$ for HCBs on a $16\times 16$ checkerboard lattice with $\Delta_1=16$, $\Delta_2=10$ and $\Delta_3=5$. The figure demonstrates the existence of supersolidity in the vicinity of $m=0.375$. The results are obtained by averaging over simulations for three different random number seeds.}
\label{16x16_result}
\end{figure}
\begin{figure}[t]
 \centering
  \begin{minipage}[c]{0.225\textwidth}
        \includegraphics[width=\textwidth]{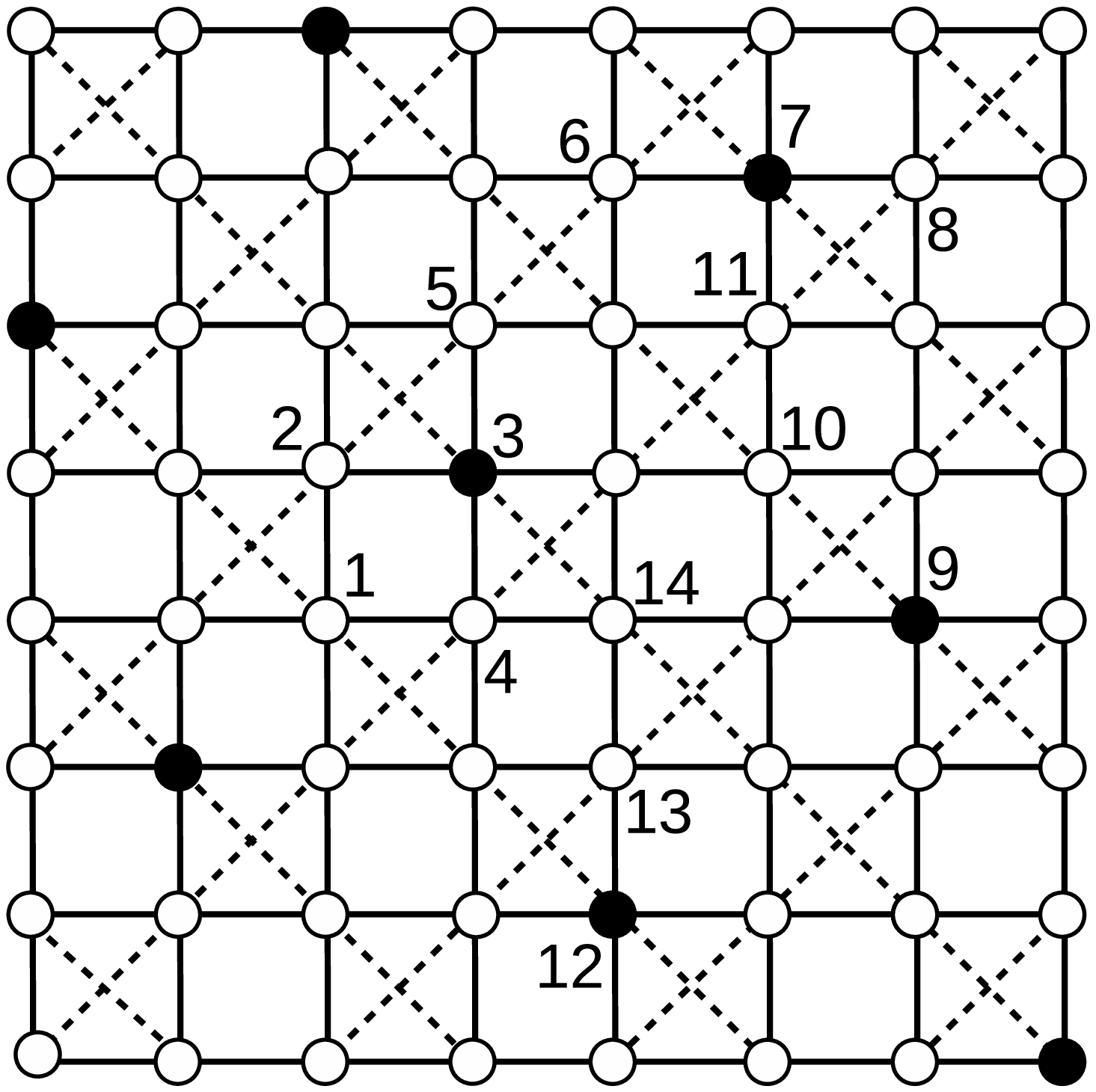}
            \vskip 1ex  (a)
            \end{minipage}
            \quad
      \begin{minipage}[c]{0.225\textwidth}
        \includegraphics[width=\textwidth]{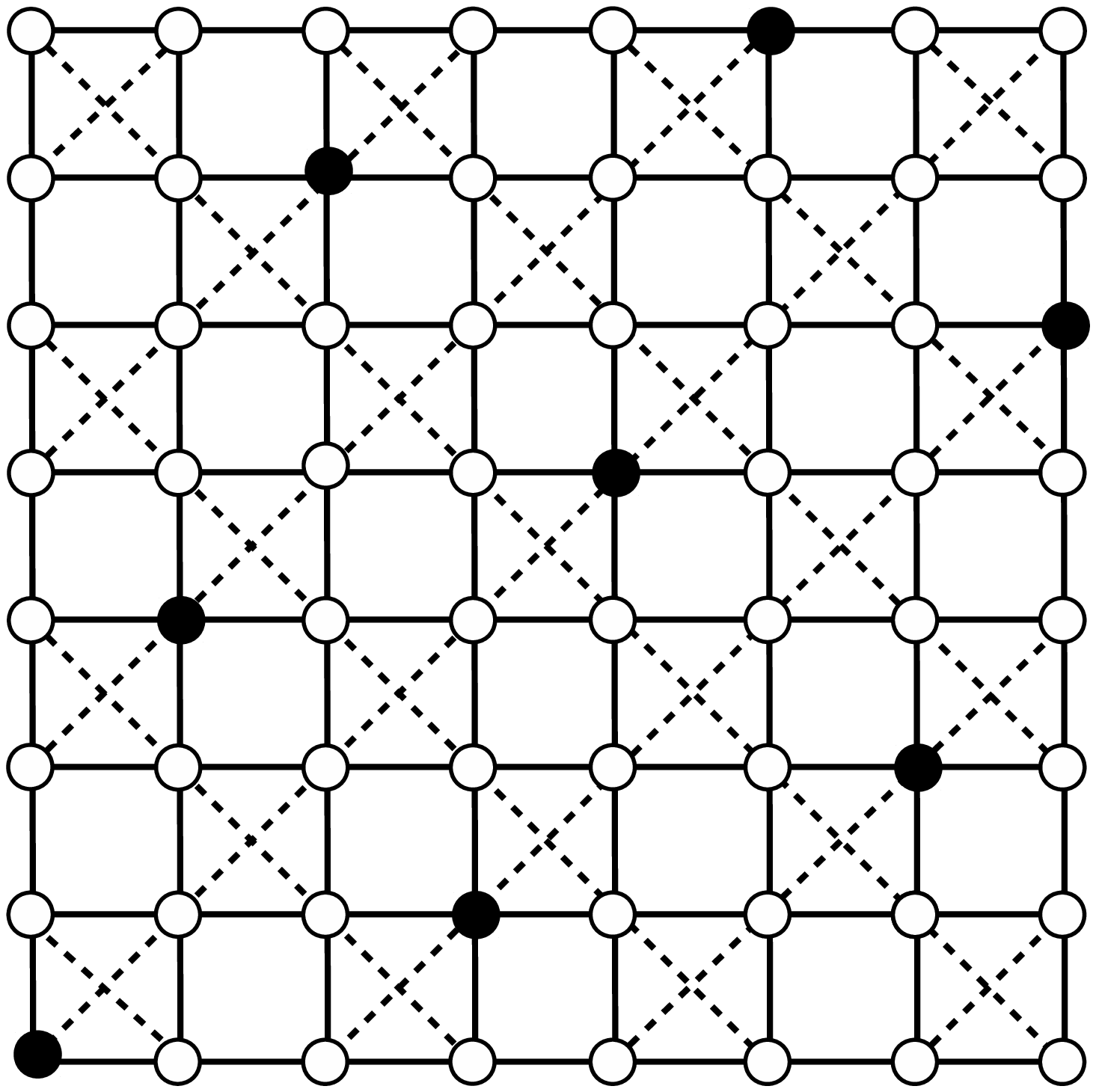}
               \vskip 1ex (b)
      \end{minipage} 
      \begin{minipage}[c]{0.225\textwidth}
        \includegraphics[width=\textwidth]{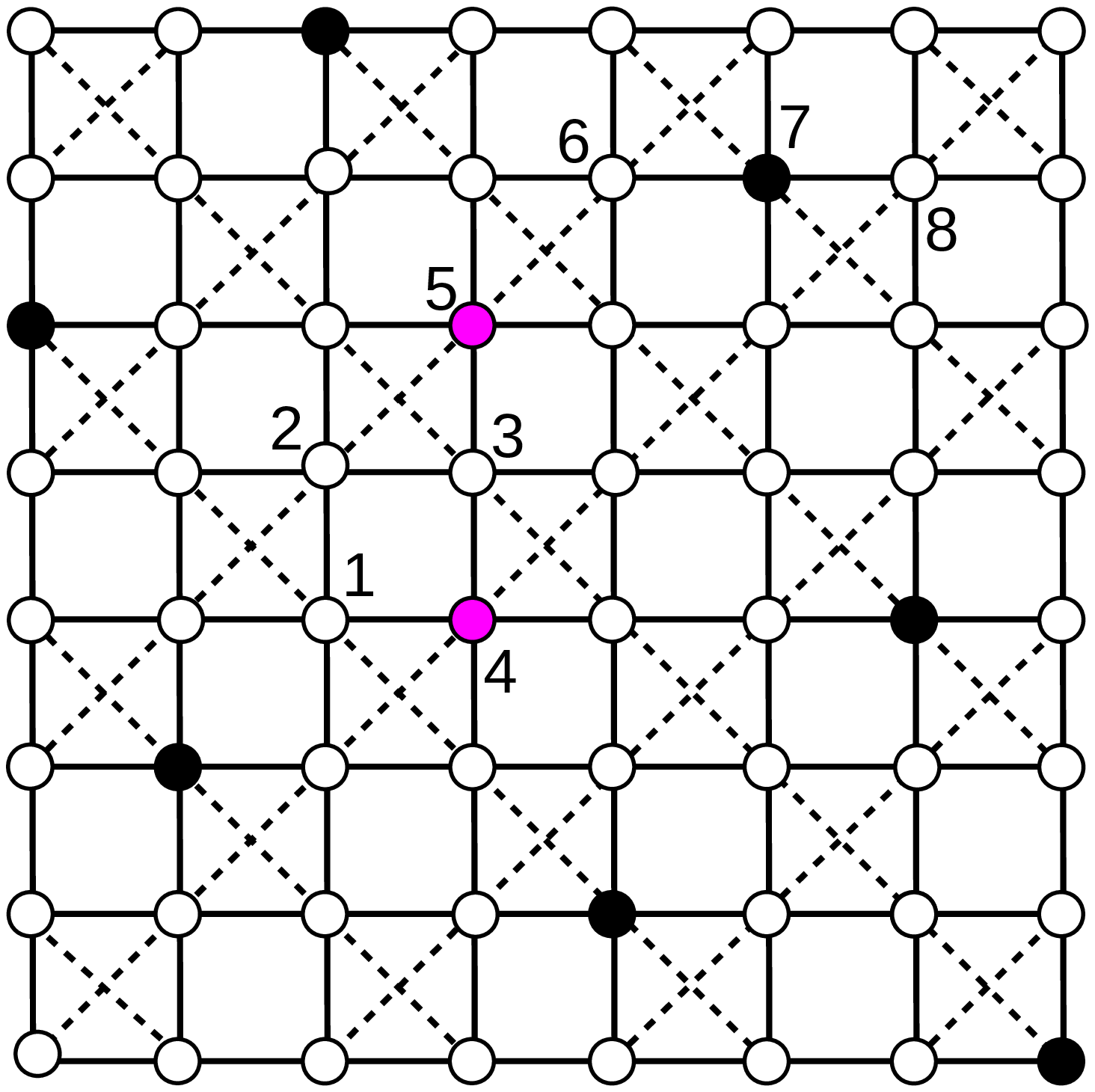}
            \vskip 1ex  (c)
            \end{minipage}
            \quad
      \begin{minipage}[c]{0.225\textwidth}
        \includegraphics[width=\textwidth]{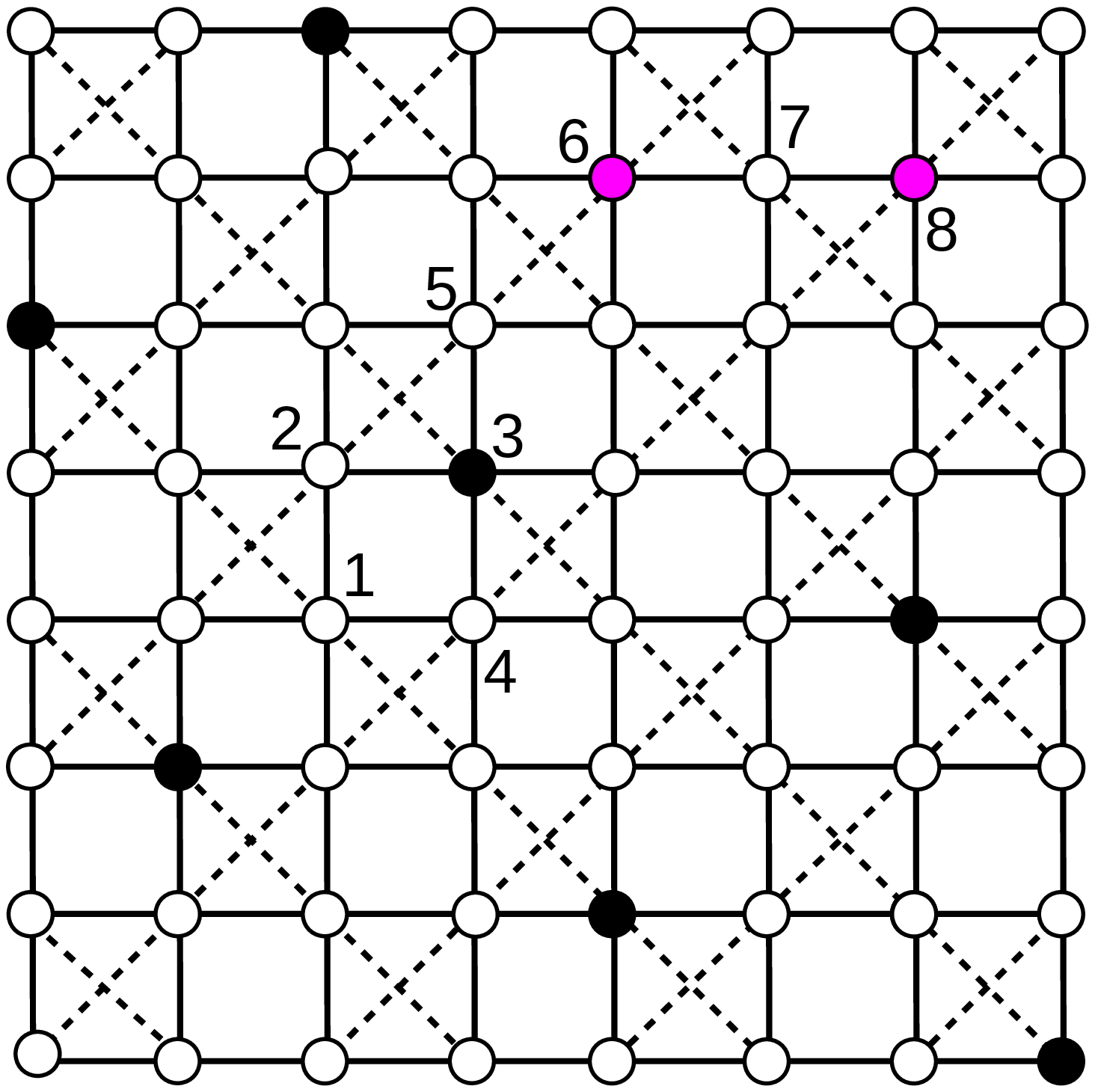}
               \vskip 1ex (d)
      \end{minipage}  
 \caption{Two different types of CDWs: (a) diagonal striped solid (dsS) indicated by a peak in the structure factor 
 at wavevector
    {$\vec{q}_1=(\pi/4,3\pi/4)$}; (b) dsS characterized by ordering wavevector {$\vec{q}_2 = (\pi/4,5\pi/4)$}. (c) A minimum energy configuration obtained after rearrangement when an extra HCB is added at site $1$ in Fig. 4(a). The rearranged particles are indicated in magenta. (d) A resulting configuration when the pair of HCBs at sites $4$ and $5$ in Fig. 4(c) flows through the system.
       } \label{fig:CDW}   
\end{figure}

To determine the various phases of the two dimensional Hubbard-Holstein model, we first set the cutoff values of the anisotropies to be $\Delta_1=16$, $\Delta_2=10$ and $\Delta_3=5$; we calculate the order parameters for magnetization $m$ values ranging from $0.25$ to $0.5$. 
The requirement that $\Delta_1\rightarrow\infty$ is implemented via a suitable choice of large but finite value of $\Delta_1$  
so as to avoid computational problems. 
Fig. \ref{16x16_result} shows the variation of the structure factor $S(\vec{Q})$ and superfluid density $\rho_s$ as the magnetization of the system is varied from $0.25$ to $0.5$; this corresponds to the variation of filling fraction of HCBs from $3/4$ to $1$. Due to the particle-hole symmetry of the Hamiltonian, the physics at filling fraction $3/4$ is the same as the one revealed at filling fraction $1/4$. Hence, in the text, we use them interchangeably at our convenience. From Fig. \ref{16x16_result},  at filling $7/8$ (i.e., $m=3/8$), we see that the system manifests a CDW state, whereas the superfluid (SF) order ceases to exist. At  filling fraction $1/8$, the HCBs arrange themselves so that no repulsion is felt; the resulting state is an insulating CDW,
characterized by a peak in the structure factor at wavevectors {$\vec{q}_1=(\pi/4,3\pi/4)$} [as shown in Fig. \ref{fig:CDW}(a)] or {$\vec{q}_2=(\pi/4,5\pi/4)$}
[as depicted in Fig. \ref{fig:CDW}(b)]. We call this CDW state a diagonal striped solid (dsS).
{One should note that, unlike the well-known checkerboard solid 
identified by the peak in the structure factor $S(\pi,\pi)$ (see Figs. 4 and 5 of Ref. \onlinecite{ghosh2}), a single wavevector is inadequate to characterize  the two equally probable CDW states at  filling $1/8$. Whenever the system manifests a dsS equivalent to that in Fig. \ref{fig:CDW}(a), $S(\vec{q}_1)$ acquires a non-zero value while  $S(\vec{q}_2)$ concomitatntly vanishes. On the other hand, for a dsS corresponding to Fig. \ref{fig:CDW}(b), the situation is reversed with $S(\vec{q}_2)$  taking a non-zero value whereas  $S(\vec{q}_1)$ now vanishing. Therefore, to identify the insulating dsS at  filling $1/8$, we should plot the sum ($S(\vec{q}_1)+S(\vec{q}_2)$) of these two structure factors.}

\begin{figure}[t]
\includegraphics[width=0.9\linewidth,angle=0]{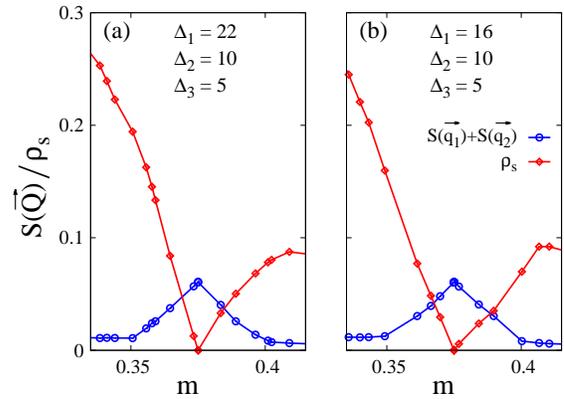}
\caption{(Color online) Comparison of the behavior of the order parameters, structure factor $S(\vec{Q})$ and superfluid density $\rho_s$, as functions of magnetization $m$ on an $8\times 8$ checkerboard lattice for two different sets of anisotropy values: (a) $\Delta_1=22$, $\Delta_2=10$, $\Delta_3=5$ and (b) $\Delta_1=16$, $\Delta_2=10$, $\Delta_3=5$. The figures demonstrate that the essential coexistence features are not altered much when $\Delta_1$ is increased beyond $16$.}
\label{compare}
\end{figure}

Now, when we add one extra particle to the system at filling $1/8$, one would normally think of two different possible scenarios. The extra particle can either occupy any empty site along the half-filled stripes or an empty one between any two stripes. First, let us assume that the particle occupies site 1 (i.e., a site along one of the half-filled stripes) in Fig. \ref{fig:CDW}(a). The repulsion felt by this particle is $2V_2+2V_3=3V_2$.  Instead of this configuration, if the particle at site 3 is moved to site 5 and the extra particle occupies site 4, the resulting configuration [see Fig. \ref{fig:CDW}(c)] is energetically favored because the repulsion felt in this case is $5V_3=2.5V_2$. Now, by the following third-order process superflow of particles can take place in the system given by Fig. \ref{fig:CDW}(c).
First, the particle at site 7 can hop to site 8 which increases the energy of the system by $ V_3$. Next, the particle at site 5 can hop to site 6 with the energy of the system being the same as that after the first process. Finally, the particle at site 4 can hop to site 3 resulting in the configuration depicted 
in Fig. \ref{fig:CDW}(d). The energy of this final configuration is the same as that of the starting configuration  shown in \ref{fig:CDW}(c). The energy of this third-order perturbation process is thus proportional to ${t_1}^3/V_3^2$. There may also be other processes by which the system manifests supersolidity when
the dsS is doped with particles. Nevertheless, this particular process is one of the possible mechanisms which gives rise to supersolidity on the interstitial side of the CDW at  filling $1/8$.
\begin{figure}[b]
\includegraphics[width=0.9\linewidth,angle=0]{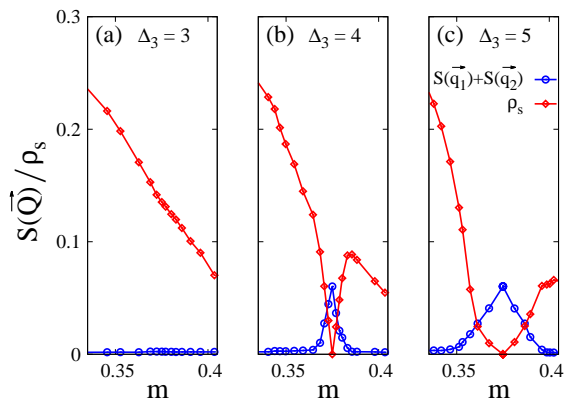}
\caption{(Color online) Variation of structure factor $S(\vec{Q})$ and superfluid density $\rho_s$ as functions of magnetization $m$ in the vicinity of filling fraction $1/8$ at three different values of $\Delta_3$ (with $\Delta_2=2\Delta_3$) and for a fixed $\Delta_1=16$: (a) $\Delta_3=3$; (b) $\Delta_3=4$; and (c) $\Delta_3=5$. The figures depict evolution of supersolidity around $m=0.375$.}
\label{diff_V2}
\end{figure}

Next, in the second possible scenario (where the extra particle occupies an empty site between any two half-filled stripes), let us assume 
that the extra particle
occupies site 2 in Fig. \ref{fig:CDW}(a). Then, the repulsion felt by this particle is $V_1+2V_3=V_1+V_2$. It is important to note  that in this case
there is no way to avoid NN occupation of HCBs; this is not allowed because two singlets cannot share an electron. Although the cutoff values of the repulsions used in our simulation makes the first scenario energetically favorable, the energy difference between these two situations is marginal. Moreover, in the second case the extra particle can hop to any of its unoccupied NN sites leading to lower energy  and eventually to supersolidity. Therefore, unless the energy difference between these two scenarios is reasonable we can not rule out the possibility of the second one. As mentioned earlier, numerical restrictions do not 
allow us to use anisotropies larger than the cutoff values used in our simulations on $16\times16$ lattices. Therefore, to avoid prohibitively large
simulation times, we considered a smaller
$8\times 8$ system and calculated the order parameters for two different sets of parameters: $\Delta_1=22$, $\Delta_2=10$, and $\Delta_3=5$;  $\Delta_1=16$, $\Delta_2=10$, and $\Delta_3=5$. Fig. \ref{compare} compares the plots of the structure factor $S(\vec{Q})$ and superfluid density $\rho_s$ as a function of magnetization $m$ for these two different sets of anisotropies. For the first set of parameters (i.e., $\Delta_1=22$, $\Delta_2=10$, and $\Delta_3=5$), 
the energy of the system with an additional particle at site 2 in Fig. \ref{fig:CDW}(a) (with $V_1+V_2=64t_1$) is much larger than the energy corresponding to
the situation in Fig. \ref{fig:CDW}(c) (with $2.5V_2=50t_1$). Hence,  we can definitely rule out the possibility of the second scenario involving the extra particle  occupying any empty site between any two half-filled stripes. Since  Fig. \ref{compare} demonstrates that both the parameter sets yield similar results,  we can capture the essential physics of the two-dimensional Hubbard-Holstein model by using $\Delta_1=16$, $\Delta_2=10$ and $\Delta_3=5$ as the cutoff values of the anisotropies in the simulations. 
{It should be noted that, in Fig. \ref{compare}, the non-zero value of the structure factor $S(\vec{Q})$, below $m\approx0.35$ and beyond $m\approx0.4$, is
just an artifact of the small system size.}
\begin{figure}[t]
\includegraphics[width=0.9\linewidth,angle=0]{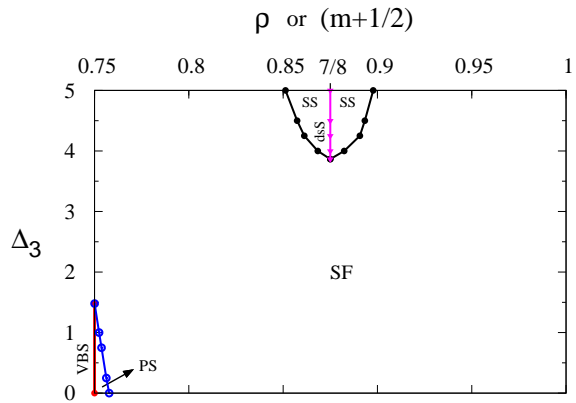}
\caption{(Color online) Ground state phase diagram in terms of filling fraction $\rho$ (or magnetization $m$) for HCBs on a $16\times 16$ checkerboard lattice. Here dsS represents diagonal striped solid, SS stands for the supersolid phase corresponding to dsS, VBS denotes valence-bond solid and PS represents the phase-separated region.}
\label{phase_diagram}
\end{figure}

As regards the vacancy side  of the half-filled diagonal striped phase, the mechanism responsible for supersolidity can be explained as follows. Let us assume that we remove two HCBs from sites 3 and 7 in the configuration depicted in Fig. \ref{fig:CDW}(a). Then, the HCB at site 9 can hop to site 10 without altering the potential energy of the system; next, this HCB at site 10 can hop to site 7 by hopping via site 11 and again the overall potential energy of the system  remains unaltered at the end of each hopping process. Similarly, again through a three-step hopping process, without any additional potential energy cost, the particle at site 12 can hop to site 3 by sequentially hopping through sites 13 and 14. Effectively,  the pair of holes at sites 3 and 7,  moves from one stripe to another one (where they occupy sites 12 and 9),  and thereby  the coexistence of superfluidity and CDW is manifested.
\begin{figure}[t]
\includegraphics[width=0.9\linewidth,angle=0]{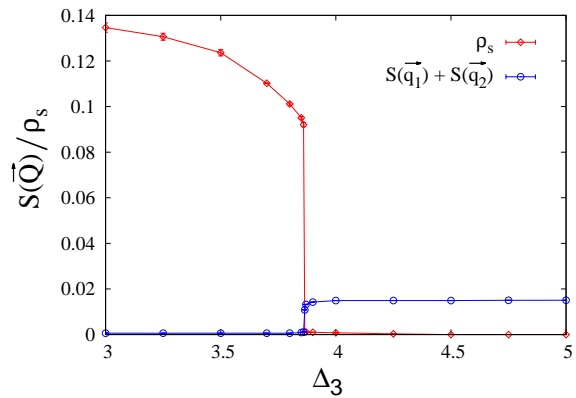}
\caption{(Color online) Plots of the order parameters, structure factor $S(\vec{Q})$ and superfluid density $\rho_s$, as functions of the NNNN anisotropy $\Delta_3$ at magnetization $m=0.375$ (corresponding to $7/8$ filling), $\Delta_2=2\Delta_3$ and $\Delta_1=16$. First order transition is depicted through jumps in both order parameters at $\Delta_3\approx 3.865$.}
\label{1st_order}
\end{figure}

Next, we perform a general study of the supersolid phase as a function of NNNN anisotropy $\Delta_3$, at a fixed value of $\Delta_1=16$. We vary $\Delta_3$ (with $\Delta_2=2\Delta_3$) and calculate the order parameters for magnetization values ranging from $0.25$ to $0.5$. Fig. \ref{diff_V2} displays the variation of the structure factor $S(\vec{Q})$ and the superfluid density $\rho_s$ as the magnetization of the system is varied in the vicinity of filling fraction $1/8$ for 
three different values of $\Delta_3$. For $\Delta_3=3$ there is no signature of any CDW at the filling $1/8$,
instead only superfluidity exists [as demonstrated in Fig. \ref{diff_V2}(a)]. Fig. \ref{diff_V2}(b) shows that, as we increase the $\Delta_3$ value to $4.0$, a diagonal striped solid (dsS) appears at HCB density $\rho=1/8$ and a supersolid (SS) region, of small width, grows on both sides of the CDW. As the NNNN anisotropy is increased further to $\Delta_3=5$, the width of the supersolid region increases further.

The ground state phase diagram is displayed in Fig. \ref{phase_diagram} for HCBs on a $16\times 16$ checkerboard lattice. At $\rho=1/8$, the system manifests the existence of a dsS when  $\Delta_3 \gtrsim 3.865$. On both sides of this CDW we have a supersolid region (SS), i.e., a homogeneous coexistence of half-filled diagonal striped solid and superfluid; further away from  $\rho=1/8$ and beyond the supersolid region, a  superfluid (SF) region exists. For $\Delta_3 \gtrsim 3.865$, as the value of NNNN anisotropy is increased, the width of the SS region increases.

In our simulations, since we can not fix the magnetization or density of the system, we tune the magnetic field $h$ to access various magnetization values. Usually, for  a fixed value of magnetic field, the resulting magnetization fluctuates during the simulation. Therefore, in the phase diagram we can not usually study the nature of the phase transition by varying the $\Delta_3$ value at a fixed magnetization (or density). However, in the  CDW state, we always have a plateau in the magnetization curve (where the magnetization of the system does not change) when plotted as a function of the magnetic field;
thus, by choosing a magnetic field in the plateau, we can ensure  a constant magnetization of the system for different values of $\Delta_3$.

For the filling $7/8$  (corresponding to $m=0.375$), as we increase the $\Delta_3$ value from $3$ to $5$, Fig. \ref{1st_order} shows that  the structure factor $S(\pi/4,3\pi/4)+S(\pi/4,5\pi/4)$ jumps dramatically from $0$ to almost its maximum value at $\Delta_3\approx3.865$; concomitantly, the superfluid density $\rho_s$ drops  to zero value. In the phase diagram, this signifies a first-order transition from a superfluid to CDW state as we move along the $\Delta_3$ axis at $m=0.375$. It is worth mentioning here 
that this transition, from a $U(1)$ symmetry broken
SF phase to a translational symmetry broken CDW phase, is consistent with Landau's picture of phase transition. An important point to note is that the magnetization can be fixed exactly at $0.375$ only after the transition to the CDW state; before the transition, i.e., in the superfluid region, the magnetization can be estimated as $m=0.375\pm 0.0001$.
\begin{figure}[t]
\includegraphics[width=0.9\linewidth,angle=0]{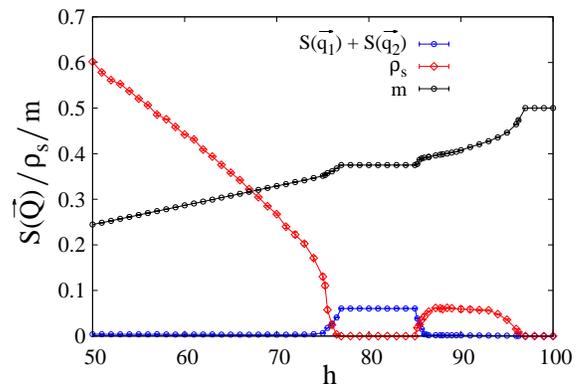}
\caption{(Color online) Variation of the order parameters (magnetization $m$, structure factor $S(\vec{Q})$, and superfluid density $\rho_s$) in terms of the magnetic field $h$ for the set of anisotropy values $\Delta_1=16$, $\Delta_2=10$, and $\Delta_3=5$. Plots depict continuous SF-SS and SS-dsS transitions.}
\label{field_vs}
\end{figure}

Next, excluding the special point ($m=0.375, \Delta_3\approx3.865$), we study the nature of the transitions along the $m$-axis of the phase diagram. Now,
Fig. \ref{diff_V2} indicates that, at a fixed value of $\Delta_3$,  the order parameters change continuously as a function of the magnetization $m$,
thereby depicting continuous phase transitions between various phases. A more reliable procedure, for detecting the nature of the phase transitions along the magnetization axis of the phase diagram, is to study the behavior of the order parameters magnetization, structure factor and superfluid density as a function of the magnetic field $h$. In Fig. \ref{field_vs}, we demonstrate that the order parameters change continuously as  the magnetic field $h$ is varied;
this  rules out the possibility of a first-order phase transition. Therefore, we conclude that all superfluid-supersolid and supersolid-solid transitions, encountered while moving along the $m$-axis of the phase diagram, are of continuous nature. { Here, it is important to note that, 
whenever there is a flat region in the magnetization curve, the superfluid density vanishes. Usually, a magnetization plateau indicates the 
presence of a gapped phase in the system\cite{IBose}. In Fig. \ref{field_vs}, the first plateau in the magnetization curve signifies the existence of the insulating CDW state dsS; consequently, the superfluid density drops down to zero. On the other hand, the second flat portion in the magnetization curve corresponds to a fully-filled system (or equivalently an empty lattice) which is not a Mott insulator. In both the cases, it is not possible for the particles to move, thus producing a zero superfluid density.}
\begin{figure}[t]
 \centering
  \begin{minipage}[c]{0.225\textwidth}
        \includegraphics[width=\textwidth]{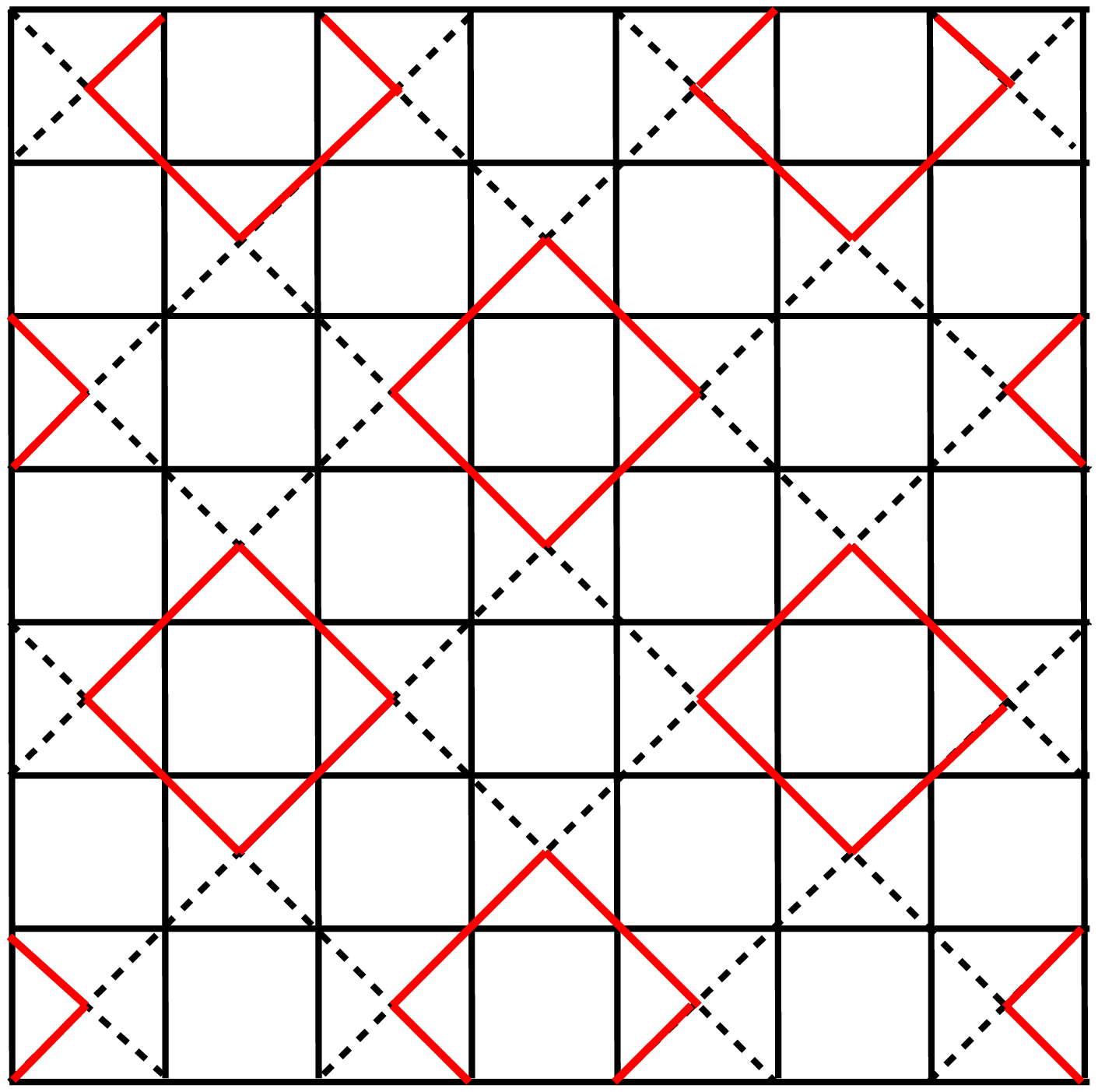}
           \vskip 1ex  (a) 
            \end{minipage}
            \quad
      \begin{minipage}[c]{0.225\textwidth}
        \includegraphics[width=\textwidth]{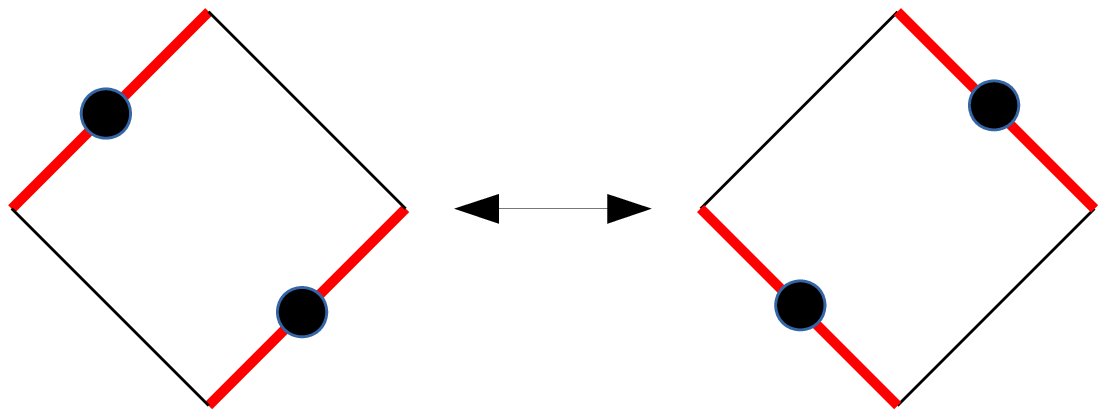}
               \vskip 1ex (b)
      \end{minipage}
      \begin{minipage}[c]{0.225\textwidth}
        \includegraphics[width=\textwidth]{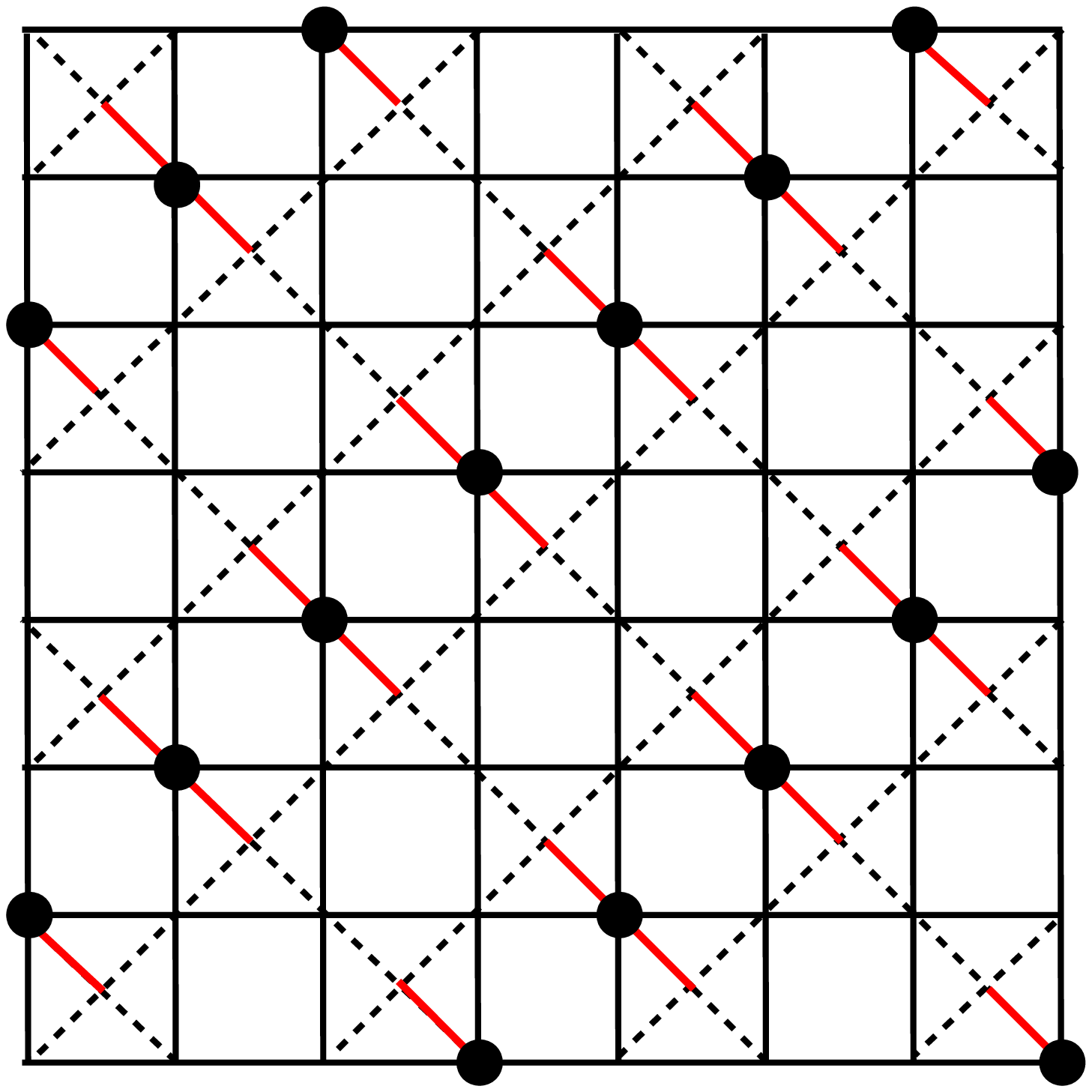}
            \vskip 1ex  (c)
            \end{minipage}
            \quad
      \begin{minipage}[c]{0.225\textwidth}
        \includegraphics[width=\textwidth]{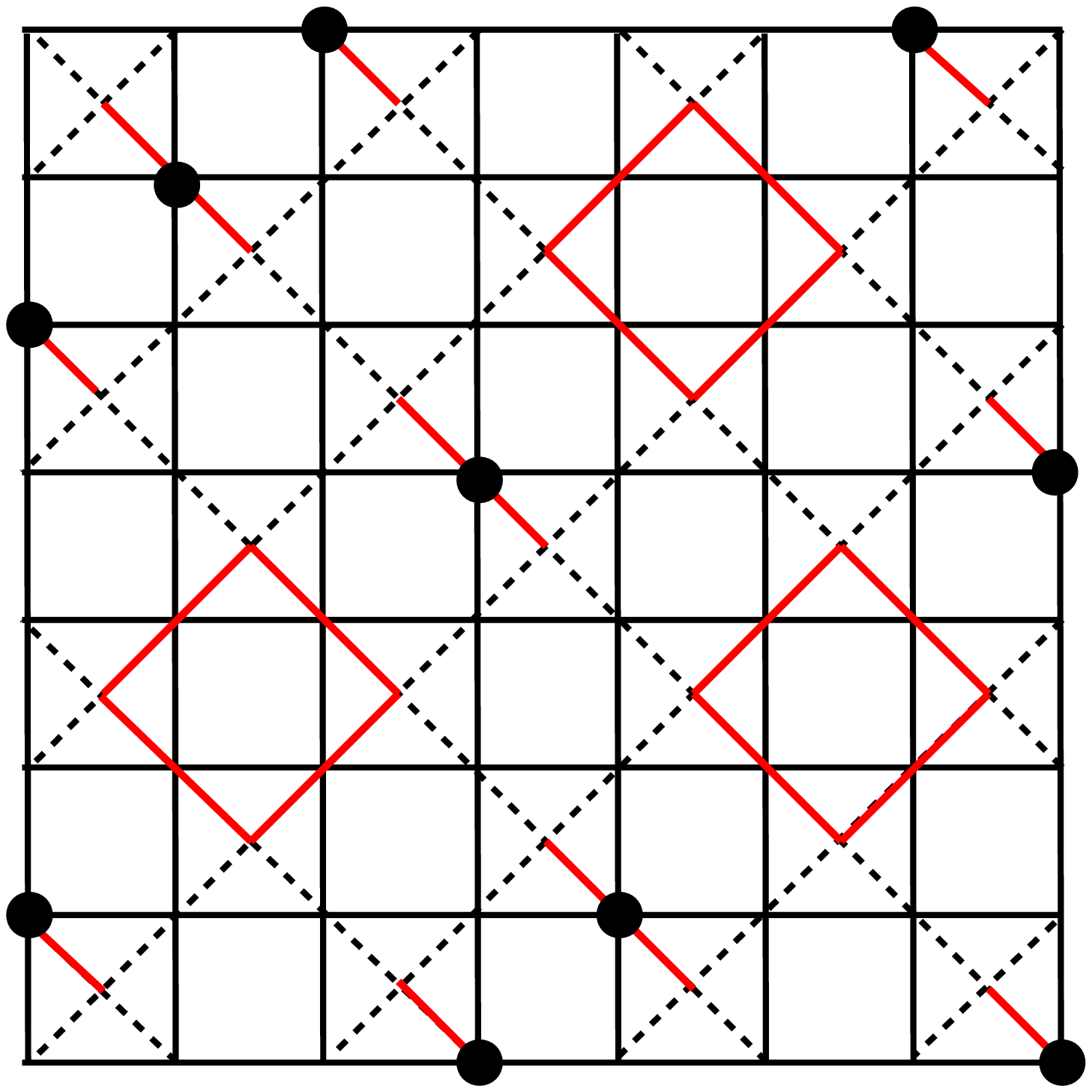}
               \vskip 1ex (d)
      \end{minipage}  
 \caption{(Color online) Different types of valence bond solids: (a) The ideal plaquette state on the checkerboard lattice where the red diamonds indicate resonance via the ring-exchange process depicted in Fig. 11(b); (c) the ideal columnar state with the black circles representing the HCBs; and (d) a mixed columnar-plaquette state.}
 \label{fig:VBS}   
\end{figure}

We now concentrate on the filling fraction $1/4$ (corresponding to $m=0.25$) in the phase diagram depicted in Fig. \ref{phase_diagram}. The quarter-filled checkerboard lattice has been studied by various authors using different types of Hamiltonians. Sen {\em et al.}\cite{Damle} and Wessel\cite{Wessel1} considered a Hamiltonian involving NN repulsion between HCBs but omitting the hopping along the diagonals of the non-void plaquettes . In Ref. \cite{Wessel2}, Wessel studied the quarter-filled checkerboard lattice using a Hamiltonian consisting of NN hopping and NN repulsion.
The study showed that beyond some particular repulsion value, the system goes through a quantum phase transition from a superfluid to an insulating valence bond solid (VBS). The  VBS can be the ideal plaquette type [shown in Fig. \ref{fig:VBS}(a)], the ideal columnar type [depicted in Fig. \ref{fig:VBS}(c)], or  a mixed columnar-plaquette phase [such as in Fig. \ref{fig:VBS}(d)]. 
\begin{figure}[t]
\includegraphics[width=0.9\linewidth,angle=0]{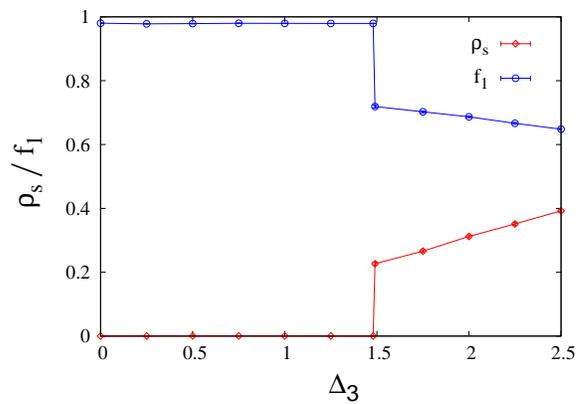}
\caption{(Color online) Plots of superfluid density $\rho_s$ and the fraction $f_1$ (representing the relative number of singly occupied non-void plaquettes) in terms of the NNNN anisotropy $\Delta_3$ at magnetization $m=0.25$ which corresponds to $1/4$ filling of HCBs in a $16\times16$ checkerboard lattice. The NN anisotropy is fixed at $\Delta_1=16$; NNN anisotropy $\Delta_2=2\Delta_3$. The figure depicts VBS-SF first-order transition at $\Delta_3\approx1.48$.}
\label{1_4_1storder}
\end{figure}
To characterize these VBS states, besides employing superfluid density $\rho_s$, different order parameters were used by various authors. A common feature among all the VBS states is that each non-void plaquette is occupied by a single HCB. Therefore, along with the superfluid density, we calculate a fraction $f_1$ which denotes the relative number of  non-void plaquettes that are occupied by a single HCB. For the VBS phases this fraction $f_1$  will have a peak value $1$, whereas for any other phase it will assume a smaller non-zero value.

\begin{figure}[t]
\includegraphics[width=0.9\linewidth,angle=0]{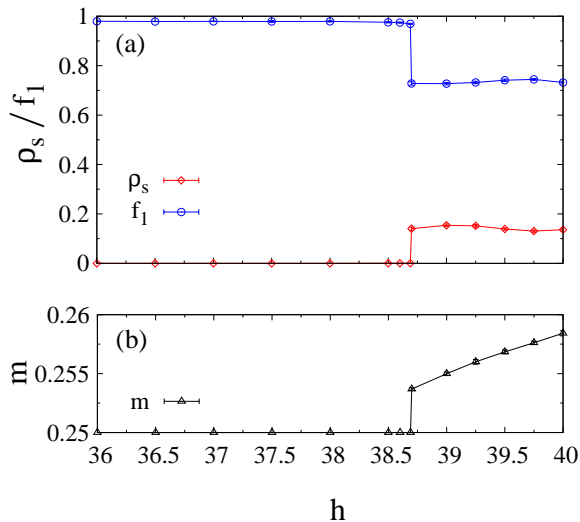}
\caption{(Color online) Plots of (a) superfluid density $\rho_s$ and fraction $f_1$ (representing the relative number of singly occupied non-void plaquettes) and (b) magnetization $m$ vs magnetic field $h$ at a fixed NNNN anisotropy $\Delta_3=0.75$ on a $16\times16$ checkerboard lattice. The NN anisotropy is again fixed at $\Delta_1=16$, while the NNN anisotropy $\Delta_2=2\Delta_3$. The sharp jump in the magnetization curve indicates the existence of a phase-separated (PS) region involving superfluid and VBS (identified by $f_1\approx1$). {Furthermore, the plateau corresponds to a Mott insulating region}.}
\label{1_4_1storder_horizontal}
\end{figure}

Fig. \ref{1_4_1storder} depicts the variation of the superfluid density $\rho_s$ and the fraction $f_1$  on a $16\times 16$ checkerboard lattice as the NNNN anisotropy $\Delta_3$ is varied from $0$ to $2.5$ (with $\Delta_2=2\Delta_3$ and $\Delta_1=16$). For lower values of $\Delta_3$, at one-fourth filling, the system manifests a VBS phase demonstrated by the close-to-unity value of
the fraction $f_1$ 
and the zero value of the superfluid density $\rho_s$. As we increase the value of $\Delta_3$, a first-order phase transition, from VBS to superfluid, is realized beyond $\Delta_3=1.48$; the transition is indicated by a jump in the superfluid density $\rho_s$ and an accompanying sudden drop in the fraction $f_1$ from its maximum value $1$ to some smaller non-zero value. In the phase diagram depicted in Fig. \ref{phase_diagram}, this signifies a first-order phase transition, along the $\Delta_3$ axis
and at $\Delta_3=1.48$, when
the magnetization remains fixed at $m=0.25$. Interestingly, while tuning $\Delta_3$,
the magnetization remains fixed at $0.25$ only in the insulating VBS phase; whereas, after the transition to the superfluid region, the magnetization
fluctuates and is estimated as $m= 0.25\pm0.000051$).
On the other hand, as magnetization is changed from $m=0.25$ while keeping $\Delta_3$ fixed, 
we can identify a phase-separated (PS) region where a jump in $m$ as well as in $f_1$ and $\rho_s$ occurs.
In Fig. \ref{1_4_1storder_horizontal}, the superfluid density $\rho_s$, the fraction $f_1$, and the magnetization $m$ of the system is shown as a function of the applied magnetic field $h$ at a fixed NNNN anisotropy $\Delta_3=0.75$. The sharp jump in the magnetization clearly manifests the existence of a PS region in the vicinity of the VBS.
The PS window in the phase diagram becomes narrower as $\Delta_3$ is increased from zero and vanishes at $\Delta_3 \approx 1.48$.
\section{Conclusions}\label{Conclusion}
In the present work, we concerned ourselves with understanding the competition and/or  cooperation of various orders within the correlated singlet phases in the 
two-dimensional Hubbard-Holstein model. 
{Strictly speaking, correlated-singlet phase requires singlets that are separated which is only possible
at fillings $\le 1/4$ in the Hubbard-Holstein model (i.e., fillings $\le 1/8$ of HCBs on a checkerboard lattice).}
Extending the results of  Refs. \onlinecite{sahinur1} and \onlinecite{sahinur2}, we arrived at the $t_1-V_1-V_2-V_3$ Hamiltonian for HCBs on a checkerboard lattice with the NN repulsion $V_1$ being infinity. We showed that the essential physics of the system can be captured
even when  cutoff values of the repulsions are used. Unlike the one-dimensional Hubbard-Holstein model, the two-dimensional  version revealed the existence of a supersolid region.   Around  filling fraction $1/8$, supersolidity is realized; whereas  at filling $1/8$, only CDW order results. This result demonstrates how the dimensionality plays an important role in stabilizing the supersolid phase. We also provide an intuitive explanation for the mechanism behind the formation of  CDW as well as the occurrence of supersolidity  on the interstitial side as well as on the vacancy side of the CDW.

Next, we performed a general study of the $t_1-V_1-V_2-V_3$ model;  by varying the NNNN repulsion $V_3$, we derived the complete phase diagram of the system in terms of the filling fraction (or magnetization) of the system. At  filling fraction $1/8$, the system reveals the existence of a half-filled diagonal striped solid. Contrastingly,  a quarter-filled system manifests the valence bond solid consistent with the literature\cite{Wessel2}. We also show that, in the phase diagram, first-order transitions are realized while going from superfluid to dsS at filling fraction $1/8$ and from VBS to superfluid phase at filling $1/4$. On the other hand, the superfluid-supersolid or the supersolid-solid transition at fixed NNNN repulsions, when we vary the magnetization of the system around
filling $1/8$, turned out 
to be of  continuous nature. Lastly, by varying the magnetization of the system around quarter filling, a PS region is identified next to the  VBS phase. 

A unique feature of our model, compared to many other models, is that the checkerboard lattice naturally emerges out of the square lattice governed by the two-dimensional Hubbard-Holstein Hamiltonian in the parameter regime where 
 correlated singlets are produced. Furthermore, unlike a number of other checkerboard models studied in the literature, the parameter values used in our model can be 
either  obtained from first-principle calculations or determined from experiments.

{Lastly, it should be emphasized that the model that we consider  (i.e., the Hubbard-Holstein model) involves a combination of  electron-electron
and electron-phonon interactions in their simplest forms. In a restricted parameter regime, this simple model is shown to manifest lattice supersolidity. Since real materials exhibiting lattice supersolidity generally involve more complexities,  further investigations are needed to figure out the relevance of our model  for such systems. 
Additionally, with the rapid advancements in artificially engineered systems, we hope that our model can be experimentally realized, thus advancing the overall understanding of different lattice-supersolid phases. }
\section{Acknowledgements}
The computing resources of the Condensed Matter Physics Division of Saha Institute of Nuclear Physics are acknowledged. S. Y. thanks P. B. Littlewood, S. Reja,
and G. Baskaran for useful discussions. A. G.  thanks M. Sarkar and S. Nag for valuable discussions during the initial stage of this work.

\end{document}